\documentclass[manuscript,screen]{acmart}
\usepackage{listings}  
\usepackage{xcolor}    

\AtBeginDocument{%
  }

\author{Chang Gong}
\affiliation{%
  \institution{School of Cyberspace Security, Hainan University}
  \city{Haikou}
  \country{China}
  \postcode{570228}
}
\authornote{Both Chang Gong and Zhongwen Li are co-first authors of the article.}
\author{Zhongwen Li}
\authornotemark[1]
\email{lizhongwen1230@gmail.com}
\affiliation{%
  \institution{School of Cyberspace Security, Hainan University}
  \city{Haikou}
  \country{China}
  \postcode{570228}
}

\author{Xiaoqi Li}
\affiliation{
  \institution{School of Cyberspace Security, Hainan University}
  \city{Haikou}
  \country{China}
  \postcode{570228}
}
\email{csxqli@ieee.org}

\begin{document}

\title{Information Security Based on LLM Approaches: A Review}



\begin{abstract}
Information security is facing increasingly severe challenges, and traditional protection means are difficult to cope with complex and changing threats. In recent years, as an emerging intelligent technology, large
language models (LLMs) have shown a broad application prospect in the field of information security. In this paper, we focus on the
key role of LLM in information security, systematically review its application progress in malicious behavior prediction, network
threat analysis, system vulnerability detection, malicious code identification, and cryptographic algorithm optimization, and explore its
potential in enhancing security protection performance. Based on neural networks and Transformer architecture, this paper analyzes
the technical basis of large language models and their advantages in natural language processing tasks. It is shown that the introduction
of large language modeling helps to improve the detection accuracy and reduce the false alarm rate of security systems. Finally, this
paper summarizes the current application results and points out that it still faces challenges in model transparency, interpretability,
and scene adaptability, among other issues. It is necessary to explore further the optimization of the model structure
and the improvement of the generalization ability to realize a more intelligent and accurate information security protection
system.
\end{abstract}



\maketitle

\section{Introduction}
\

With the rapid development of information technology, the issue of information security has become increasingly prominent, posing a common challenge for individuals, enterprises, and even countries. Threats such as leakage of personal information, loss of financial data, and network fraud are increasing; enterprises are facing hidden dangers such as database security, system vulnerability, and intellectual property protection; and at the national level, cyberattacks and information wars are seriously jeopardizing the security of critical infrastructure and information systems \cite{shen2007survey}. Therefore, improving the security protection capability of information systems has become a crucial issue that needs to be addressed urgently\cite{10.1145/3641289}.
In recent years, as artificial intelligence, particularly Large Language Models (LLMs), has made significant progress in the field of natural language processing, representative models such as GPT and BERT have been widely utilized in tasks including text generation, translation, and analysis. Its powerful language understanding and generation capabilities have also shown broad application prospects in the field of information security, such as malicious code recognition, network attack detection, and vulnerability analysis\cite{li2024stateguard}. At the same time, the safe and effective application of large language models in the field of information security has become a new direction of study \cite{siponen2007review}.

The purpose of this paper is to systematically explore the current status of the application of large language models in information security, the challenges it faces, and its practical effects\cite{al2024examining}. Through the combination of literature review, case study, and empirical research, we assess the advantages and shortcomings of the Large Language Model in enhancing information security capabilities and provide theoretical support and practical guidance for its further development in the field of information security\cite{10.1145/3744746}.
\vspace{-13pt}
\section{Background}
\subsection{Neural Network}
\

\begin{figure}[H]
    \centering
    \includegraphics[width=0.4\linewidth]{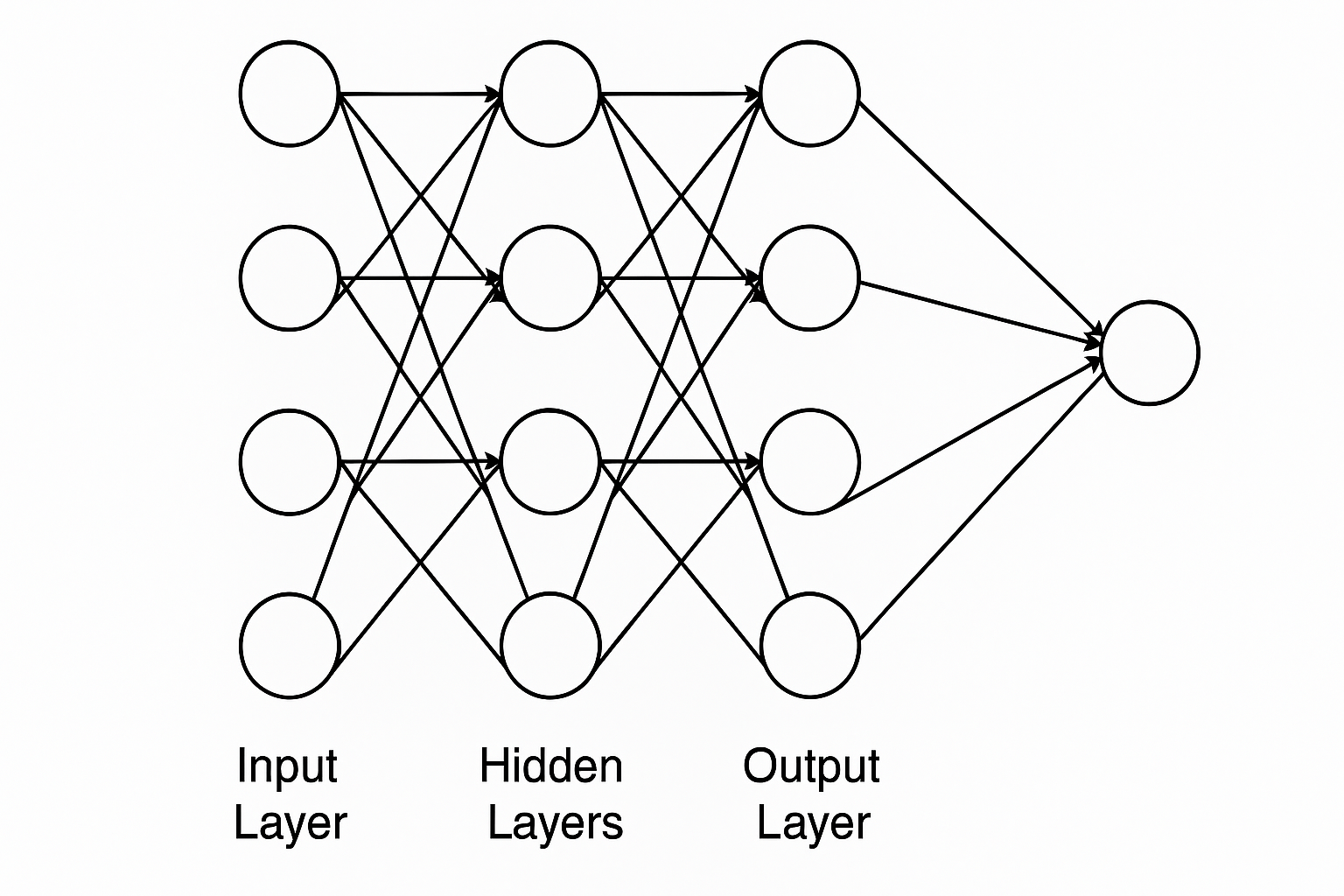}
    \caption{Neural Network Structure Diagram}
    \label{fig:1}
\end{figure}

In the field of artificial intelligence, a neural network is a mathematical model that simulates the information transfer mechanism of human neurons. Its basic structure includes an input layer, a hidden layer, and an output layer. The input layer is used to receive the original data and pass it to the subsequent network; the output layer generates the final prediction results\cite{zhao2024review}; the hidden layer is located between the two, responsible for nonlinear mapping and feature extraction of the data, and the number of layers and nodes can be flexibly adjusted according to the task requirements\cite{zhao2024improved}; generally speaking, the deeper the hidden layer is, the stronger the model's expressive ability is. Meanwhile, the training process of neural networks mainly includes two phases: forward propagation and back propagation\cite{li2024defitail}. In the forward propagation process, the input data passes through the neurons of each layer in turn, and is processed by the weighted summation and activation function to generate the final output; in the back propagation stage, the error between the prediction result and the real label is calculated, and the algorithms such as gradient descent are used to adjust the network weights in the reverse directparticularlytimize the model performance continuously. Costing multiple iterations of training, the model error is generally reduced until convergence\cite{liu2017survey}. The structure of the neural network is shown in Fig. \ref{fig:1}
\subsection{Machine Learning}
\begin{figure}[H]
    \centering
    \includegraphics[width=0.3\linewidth]{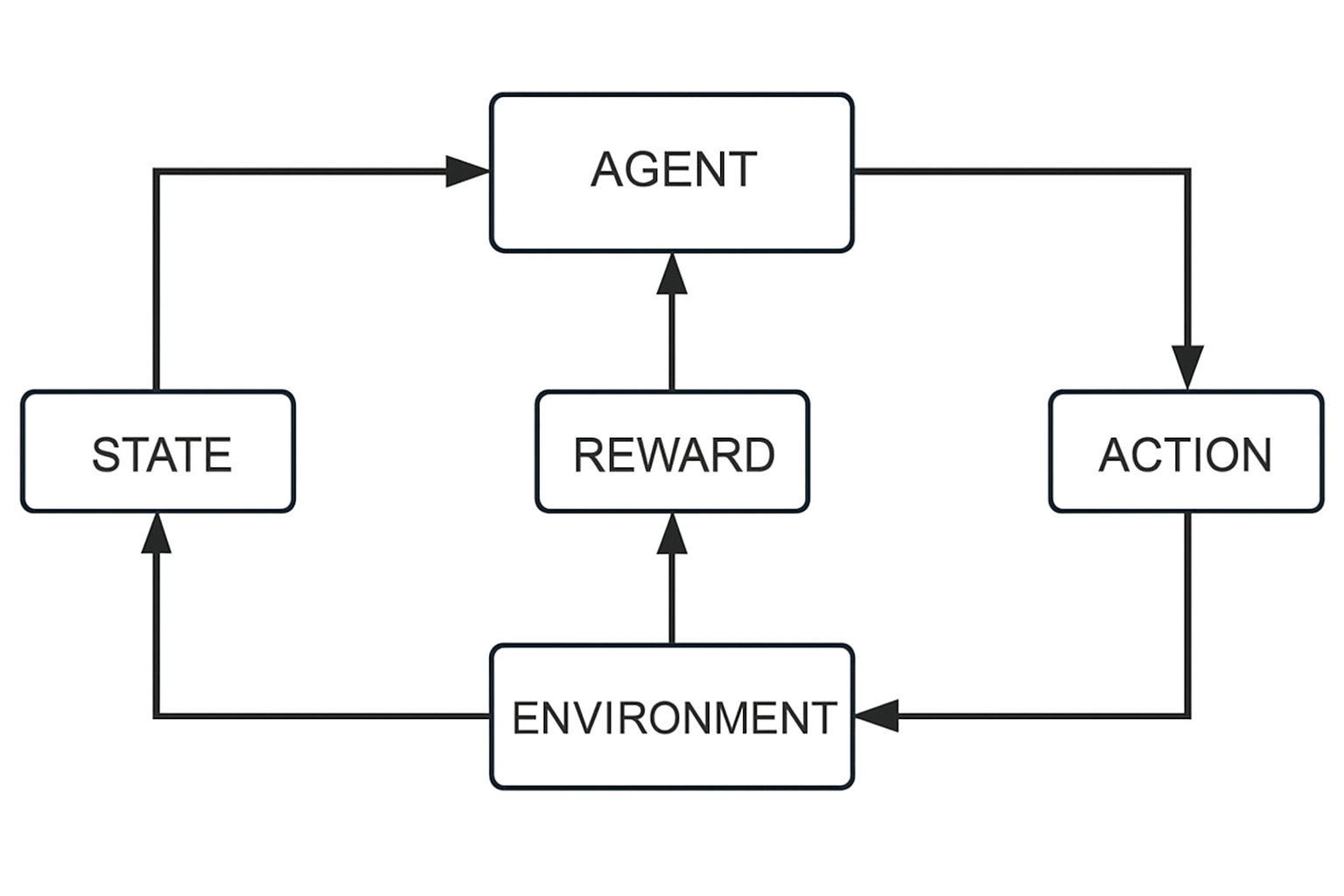}
    \caption{Reinforcement Learning Structure Diagram}
    \label{fig:2}
\end{figure}

\
\vspace{-14pt}
\begin{itemize}

    \item \textbf{Supervised Learning}: Supervised learning relies on labeled datasets for training, i.e., each sample contains known inputs with corresponding output targets\cite{igual2024supervised}. The model gradually optimizes the parameters and improves the prediction accuracy by continuously comparing the prediction results with the real labels. The method is widely used in scenarios such as spam recognition and credit scoring. However, the limitation of supervised learning is the high cost of acquiring labeled data, which relies on a large amount of manual labeling\cite{lim2024ssl}.
    \item \textbf{Unsupervised Learning}: Unsupervised learning deals with unlabeled data to discover underlying structures or patterns in the data, such as clustering and dimensionality reduction. The model automatically recognizes the intrinsic distributional features of the data without external guidance, and is commonly used for tasks such as customer segmentation and anomaly detection\cite{liu2025sok}.
    \item \textbf{Semi-supervised Learning}: Semi-supervised learning combines a small amount of labeled data with a large amount of unlabeled data for training, taking into account the advantages of supervised and unsupervised methods\cite{li2024guardians}. In practical applications, it is particularly suitable for scenarios with high labeling costs but a large amount of data, which effectively improves the model's generalization ability and efficiency.
    \item \textbf{Reinforcement Learning}: Reinforcement learning is centered on the interaction of an intelligent body (agent) with the environment to maximize long-term rewards through trial-and-error mechanisms. The agent performs actions based on its current state, receives feedback (in the form of reward or punishment) from the environment, and adjusts its strategy accordingly to improve overall performance\cite{gu2024review}. This approach is widely used in the fields of automatic control, game strategy, and robot navigation, and is one of the important paths to realize general artificial intelligence. As shown in Fig. \ref{fig:2}, the basic framework of reinforcement learning includes the key elements of state, action, strategy, and reward\cite{zhou2018brief}.
    \item \textbf{Deep Learning}: Deep learning is a machine learning method based on multi-layer neural networks, capable of hierarchical feature extraction and abstract representation of data. It has achieved remarkable results in the fields of image recognition, speech processing, and natural language understanding\cite{herrmann2024deep}. The construction of large language models is highly dependent on deep learning, and by training on large-scale corpora, the models can capture complex linguistic structure and semantic information, and realize tasks such as syntactic analysis, context understanding, and text generation\cite{chen2026deep}. In addition, deep learning also provides fundamental support for the application of large language models in information security. By training neural networks to identify potential threat patterns, the model can be used for security tasks such as malicious behavior detection, abnormal traffic identification, and data leakage prevention and control, showing significant potential in enhancing cyberspace governance capabilities\cite{pouyanfar2018survey}.
\end{itemize}

\subsection{Large Model}
\

With the continuous advancement of deep learning and computational resources, large model technology has achieved breakthroughs in the fields of natural language processing, computer vision, and speech recognition\cite{zhu2024survey}. Its development began with early artificial neural networks and back propagation algorithms, and the deep belief network proposed by Hinton in 2006 promoted the rise of deep learning, while the success of AlexNet in 2012 triggered the widespread application of deep learning. Subsequently, visual models such as VGG and ResNet have continuously optimized the network structure and improved the model performance\cite{chen2018system}. In the field of natural language processing, pre-trained language models based on the Transformer architecture, such as BERT and GPT, have been introduced one after another, promoting a leap in language comprehension and generation capabilities. Meanwhile, systems such as AlphaGo, which incorporate multiple technologies, have also verified the potential of large models in complex intelligence tasks. Currently, the development of large models is supported by key technologies such as pre-training and fine-tuning mechanisms, distributed training, model compression, and Generative Adversarial Networks (GANs), which have become an important engine for the development of AI\cite{zhao2023survey}.

Large-scale pretrained language models (PLMs) are natural language understanding and generation models trained by deep learning methods using large-scale unlabeled corpora. These models usually adopt a “pre-training + fine-tuning” strategy: the pre-training phase learns linguistic features on general-purpose text, while the fine-tuning phase optimizes on labeled data according to specific tasks. The common structure is based on the Transformer framework, which consists of two main parts: the Encoder and the Decoder, which are used to model the contextual relationship between input and output sequences\cite{zou2025malicious}. The current mainstream models can be roughly categorized into two types: autoregressive models (e.g., GPT), which predict the next word through the sequence antecedent, and self-encoding models (e.g., BERT), which use a masking mechanism to capture bidirectional contextual semantics. The former emphasizes text generation capabilities, while the latter is more suitable for comprehension-type tasks. Model training often combines a cross-entropy loss function with multi-task learning (e.g., MLM, NSP), designed to enhance semantic expressiveness and migration generalization. Pre-trained language models have been widely used in Q\&A systems, text categorization, code generation, and information security scenarios, and have become the core supporting technology in the current field of natural language processing\cite{bharathi2024analysis}.

Language modeling based on autoregressive models: Autoregressive language models use the previous known words in the sequence as conditions to predict the next word step by step, and the whole generation process satisfies the Markov property. Typical representatives include RNNLM (Recurrent Neural Network Language Model) and Transformer-based GPT series\cite{niu2024unveiling}. In these models, the hidden state of each time step is determined by the information of the previous word, and the output is calculated by a Softmax layer to predict the probability. In the case of RNNLM, the training objective is to maximize the log-likelihood of each word in the entire corpus, which is expressed mathematically as follows:
\[
P(w_1, w_2, \ldots, w_T) = \prod_{t=1}^{T} P(w_t \mid w_1, \ldots, w_{t-1})
\]
Included among these, $P(w_t \mid w_1, \ldots, w_{t-1})$ denotes the probability of the current word under the given historical conditions. In contrast, GPT models adopt a multi-layer Transformer decoder structure and utilize a self-attention mechanism to effectively capture long-distance dependencies, thus achieving stronger context modeling capabilities. Such models perform well in tasks such as text generation and code auto-completion\cite{zhou2024large}.

Self-Encoder Based Language Modeling: Unlike autoregression-based language models, autoencoder language models do not rely on conditional generation during training, but encode and decode the entire sentence as a whole. Typical models include Autoencoder Language Model (AELM) and BERT (Bidirectional Encoder Representations from Transformers).
Taking AELM as an example, the model maps the input sequence to a low-dimensional potential space and recovers the output sequence from it. During training, the model reconstructs the original input through compression and decompression operations to minimize the reconstruction error. This error is usually measured using cross-entropy, which is defined as follows:
\begin{equation}
\mathcal{L}_{\text{CE}} = - \sum_{t=1}^{T} \sum_{i=1}^{V} y_{t,i} \log(\hat{y}_{t,i})
\end{equation}
where $T$ denotes the length of the input sequence, $V$ denotes the size of the vocabulary list, $y_{t, i}$ is the actual label (one-hot encoding) of the word $i$ in the $t$ time step, and $\hat{y}_{t, i}$ is the corresponding probability predicted by the model.
The method can effectively capture the overall structural information of the input sequence and performs well in a variety of natural language processing tasks, especially for semantic understanding and context modeling\cite{kasneci2023chatgpt}.

The construction of a large language model usually involves the following key steps:
\begin{enumerate}
  \item \textbf{Data preparation:} it includes preprocessing operations such as corpus cleaning, text regularization, word splitting, vectorization, etc., to provide high-quality input data for model training.
  
  \item \textbf{Model construction:} The core module for constructing language models. Commonly used structures include Convolutional Neural Networks (CNN), Recurrent Neural Networks (RNN), and self-attention mechanisms (e.g., Transformer).
  
  \item \textbf{Pre-training phase:} Pre-training on a large-scale unsupervised corpus, often using strategies such as Masked Language Modeling (MLM) or autoregressive modeling, is used to learn contextual semantics by predicting masked words.
  
  \item \textbf{Fine-tuning phase:} Supervised fine-tuning of pre-trained models on task-specific datasets to adapt them to specific application scenarios such as classification, Q\&A, and summarization. Smaller labeled datasets are usually used in this phase, and appropriate loss functions and optimization methods are selected based on the task objectives.
  
  \item \textbf{Integration and Testing:} Integrate and deploy the model, and optimize and adjust it using performance evaluation and error analysis to ensure its effectiveness and robustness in real tasks.
\end{enumerate}

\subsection{Transformer}
\

\begin{figure}[H]
    \centering
    \includegraphics[width=0.4\linewidth]{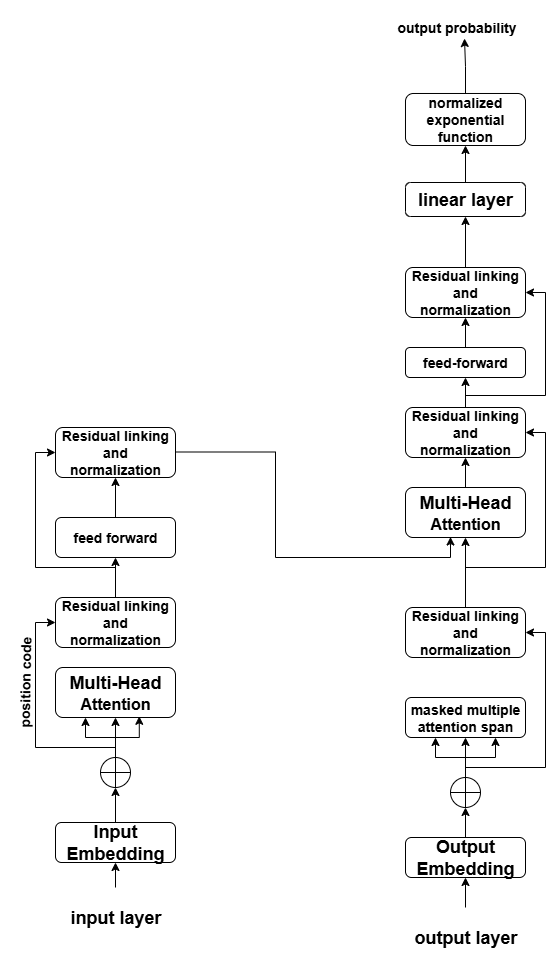}
    \caption{Transformer Overall Architecture Diagram}
    \label{fig:2-3}
\end{figure}

 To enhance the capability of natural language understanding and deep semantic modeling, the Transformer architecture has been proposed. Originally designed for machine translation tasks, the basic structure of the Transformer is shown in  Fig. \ref{fig:2-3}. The overall structure can be divided into left and right parts: the Encoder on the left side and the Decoder on the right side.
In the Encoder section, each word in the input sequence is first combined with Positional Encoding through Input Embedding to generate vector representations. These representations are then sequentially processed through multiple layers consisting of Multi-Head Attention and Feed-Forward Neural Network\cite{zhong2023tackling}, each equipped with Residual Connection and Layer Normalization ( Residual Connection and Layer Normalization) between each layer.
The decoder structure is similar to the encoder, but introduces Masked Multi-Head Attention after Multi-Head Attention to avoid future information leakage and receives the output from the encoder for inter-attention computation. Finally, after Linear Layer and Softmax (normalized exponential function) operations, the model outputs the corresponding translation or prediction results\cite{khan2022transformers}.
\subsection{Network Security Attacks}
\

Network security attack refers to the attacker's use of technical means of computer systems, network resources, or data and information for illegal access, theft, destruction, or tampering behavior\cite{hoque2014network}. According to the different attack methods and targets, common types of network attacks mainly include:
\begin{itemize}
    \item \textbf{Trojan Horse attack: } Embedding malicious code into a system by disguising it as a legitimate program to achieve remote control, information theft, or destructive operations without the user's knowledge.
    
    \item \textbf{Virus attack: } A type of self-replicating malicious code that spreads by infecting a host program and can cause file corruption, system crash, or information leakage.
    
    \item \textbf{Worm attack: } Self-replicating malicious code that does not depend on the host program, spreads mainly through the network, can infect quickly, and puts great pressure on network bandwidth and system resources\cite{bu2025smartbugbert}.
    
    \item \textbf{Distributed Denial of Service Attack (DDoS):} The use of a large number of controlled hosts to launch concurrent requests to a target, causing it to run out of resources and disrupt its services, thereby paralyzing the system.
    
    \item \textbf{SQL Injection Attacks:} Exploit the security vulnerabilities of web applications in processing user input to control the database by injecting malicious SQL statements to realize information theft or destructive operations.
    
    \item \textbf{Social Engineering: } Attacks based on human psychological weaknesses, such as forging phishing emails, fake websites, and other means to induce users to disclose sensitive information.
\end{itemize}
In the face of evolving network attack methods, there is an urgent need to take systematic security measures\cite{ehlert2010survey}, such as strengthening the authentication mechanism, regularly updating system patches, deploying Intrusion Detection Systems (IDS), and improving user security awareness. Meanwhile, attack detection technology also plays a key role in the network security defense system, and its core methods include:
\begin{itemize}
    \item \textbf{Network Traffic Analysis:} Identify anomalous communication patterns and potential attack activities by analyzing the behavior of data flows at the transport and application layers.
    
    \item \textbf{Log Analysis:} Use the collection and analysis of system and application logs to achieve rapid location and response to events such as operational anomalies and intrusion behaviors.
    
    \item \textbf{Behavioral Analysis:} Modeling and anomaly detection based on user or system operational behavioral characteristics helps to discover hidden attacks such as Advanced Persistent Threats (APTs)\cite{li2021hybrid}.
\end{itemize}
\subsection{Security Breach}
\

A security vulnerability usually refers to a security gap in a system caused by design flaws or improper configuration in the implementation of hardware, software, or protocols, whereby an attacker can access or destroy system resources without authorization, leading to information leakage, system paralysis, or service interruption. In actual operation, vulnerabilities may trigger abnormal changes in system structure or data loss, thus bringing irreversible security risks and threatening the normal use of computer systems.

Although common security tools such as antivirus software have a certain vulnerability repair function, with the continuous evolution of attack technology, new types of vulnerabilities continue to emerge, and their hidden complexity is also increasing\cite{li2017discovering}. Some vulnerabilities are hidden in the underlying structure of the system or triggered by the interaction of multiple components, which are difficult to discover and repair promptly through traditional methods. At the same time, the continuous compression of the software development cycle has also accelerated the pace from the emergence of vulnerabilities to be exploited, so that the existing detection methods face the double challenge of efficiency and accuracy\cite{li2024dark}.

In addition, vulnerability detection often relies on high-level technical experts and complex detection tools, which puts high demands on resource allocation and technical capabilities. Therefore, building an efficient, intelligent, and automated vulnerability detection mechanism has become a key direction to improve network defense capability. In recent years, Large Language Models (LLMs) have shown strong potential in code semantic understanding and vulnerability prediction, promoting the transformation of vulnerability detection technology to deep learning-driven intelligence, and providing a brand new way of thinking to improve system security and robustness\cite{al2020dealing}.
\subsection{Malicious Code}
\

Malicious code refers to malicious programs designed to disrupt system functionality or steal user data. Common types include viruses, worms, Trojan horses, and spyware. Viruses and worms can replicate themselves, with worms spreading autonomously over the network; Trojans use camouflage to lure users to execute them; and spyware focuses on surveillance behavior and information theft. According to the propagation path, malicious code can be spread through email, web pages, removable devices, and other media; according to the target of attack, the target can cover individual users, enterprise networks, and even government systems.
Classification and characterization of malicious code are the basis for building detection and defense mechanisms. Behavioral features such as file operations, registry modifications, system calls, and network communications, as well as static attributes such as string patterns and structural features, are the important basis for malicious code identification\cite{alazab2015profiling}.

Current mainstream detection methods include static analysis and dynamic analysis. The former analyzes the structure and logic of the source code or binary file to determine the potential risk without running the program, which is suitable for preventive detection but susceptible to obfuscation and encryption techniques; the latter monitors the program behavior during runtime, which is more reflective of the real threat but may be interfered with by the evasion mechanism of malicious code on the detection environment\cite{patsakis2024assessing}.
\vspace{-4ex}
\subsection{Cryptography}
\

Cryptography is the core discipline of information security, which mainly contains two aspects of cryptographic coding and cryptanalysis. Cryptographic coding typically committed to involves generating programs to protect information, encrypting plaintext with encryption algorithms decrypting ciphertext with decryption algorithms\cite{bu2025enhancing}. A complete cryptographic system usually contains plaintext, ciphertext, key, encryption algorithm, and decryption algorithm, and other basic elements; the three work together to ensure the confidentiality, integrity, and availability of information\cite{subramani2025review}.

Modern cryptosystems are divided into two categories: symmetric encryption and asymmetric encryption. Symmetric encryption uses the same key to complete the encryption and decryption process, which has the advantages of fast encryption and decryption speed and high computational efficiency, and is suitable for the protection of large amounts of data. Typical symmetric encryption algorithms include AES, DES, and so on. Asymmetric encryption is based on a pair of keys - public key (public key) and private key (private key), public key for encryption, private key for decryption. The security depends on the computational complexity of the mathematical puzzle. Common asymmetric encryption algorithms include RSA, DSA, and Elliptic Curve Cryptography (ECC)\cite{hellman2002overview}.

\begin{figure}[H]
    \centering
    \includegraphics[width=0.4\linewidth]{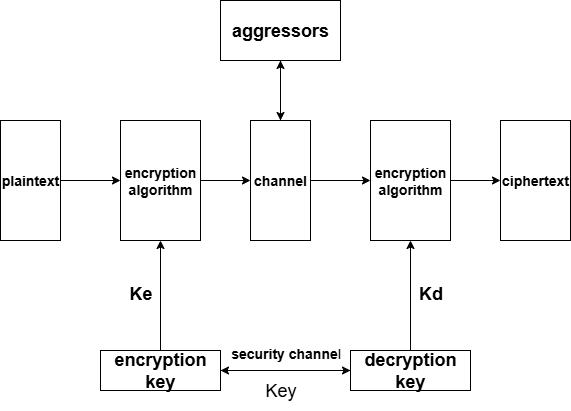}
    \caption{Modern Cryptography}
    \label{fig:6-1}
\end{figure}

Cryptographic algorithms should be designed to satisfy the correctness requirement, i.e., decrypting the encrypted ciphertext should restore the original plaintext. In addition, according to Kirchhoff's principle, the security of a cryptosystem should depend only on the secrecy of the key; even if an attacker has all the details of the encryption algorithm, the system remains reliable as long as the key is secure\cite{radanliev2024artificial}. Modern cryptographic communication typically involves generating keys through key generation algorithms, encrypting plaintext with encryption algorithms, and decrypting ciphertext with decryption algorithms, thereby constituting a complete secure transmission process. The process is shown in Fig. \ref{fig:6-1}.
\section{Malware Detection}
\

Malware behavior identification aims to detect potential threats by analyzing the operation patterns of programs (e.g., file access, registry modification, network communication, and process creation, etc.). In recent years, large language models (LLMs) have been gradually introduced into malware detection tasks due to their excellent semantic understanding and modeling capabilities\cite{bensaoud2024survey}. Compared with traditional machine learning methods that rely on a large number of labeled samples, feature engineering, and computational resources, detection methods based on large language models are able to learn complex behavioral semantic patterns directly from raw data and identify unknown malicious behaviors more efficiently. By fusing pre-trained language models with behavioral sequence modeling techniques, the method shows significant advantages in improving detection accuracy and reducing false alarm rate and becomes an important means to build an intelligent malware detection system\cite{liu2024gastrace}.

\begin{figure}[H]
    \centering
    \includegraphics[width=0.5\linewidth]{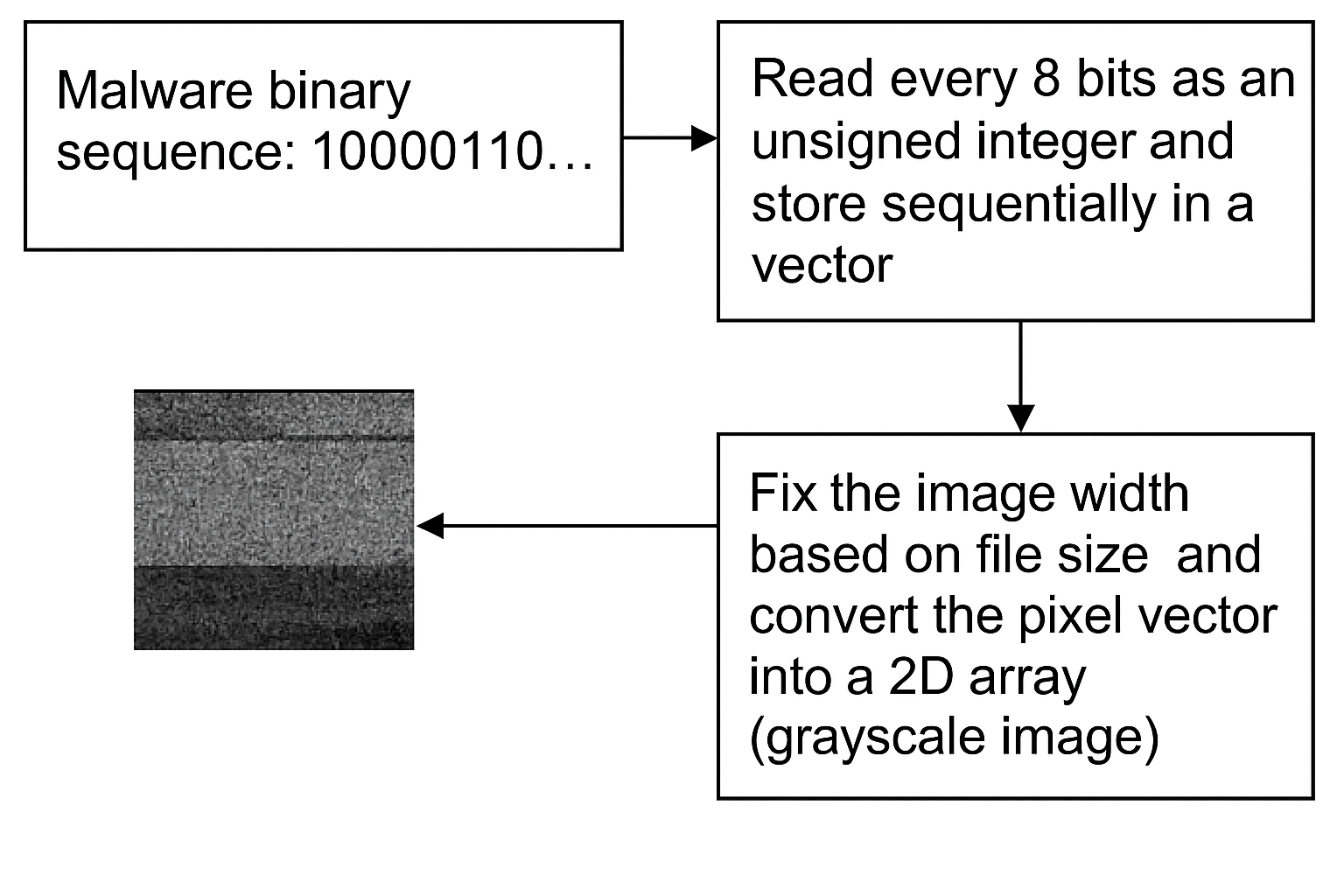}
    \caption{Binary to Malware Image}
    \label{fig:3-1}
\end{figure}

In recent years, researchers have been exploring the application of large language models and deep learning methods to malware detection and classification tasks, and have achieved remarkable results. Qiao\cite{qiao2021malware} et al. propose a malware classification method that combines Word2Vec and multilayer perceptron (MLP), and the model performs well in both traditional and IoT malware classification tasks by converting byte sequences into vectors and inputting them into a deep neural network. Victor et al. experimentally verified the potential of ChatGPT for malware identification by driving a scanner that successfully detects multiple malicious processes and avoids interference from benign processes\cite{kasneci2023chatgpt}. Alzaidy et al. applied RNNs and CNNs to classify Windows executable files and analyzed the impact of adversarial samples based on the results. Alzaidy et al. apply RNN and CNN to Windows executable file classification and analyze the impact of adversarial examples based on the results, which show that JSMA and C\&W attacks can circumvent the model detection to a certain extent, and the robustness of the attacks varies among different models\cite{alzaidy2024adversarial}. Marwaha et al. propose converting the binary files of Android malware into RGB images for visual classification using the VGG16 network. This method significantly improves classification accuracy and the F1 score, highlighting the unique advantages of visual features in malware detection\cite{marwaha2023retracted}. The binary to malware image flow is shown in Fig. \ref{fig:3-1}.

\section{Cybersecurity Analysis}
\subsection{DDoS Attack Prediction}
\

\begin{figure}[H]
    \centering
    \includegraphics[width=0.5\linewidth]{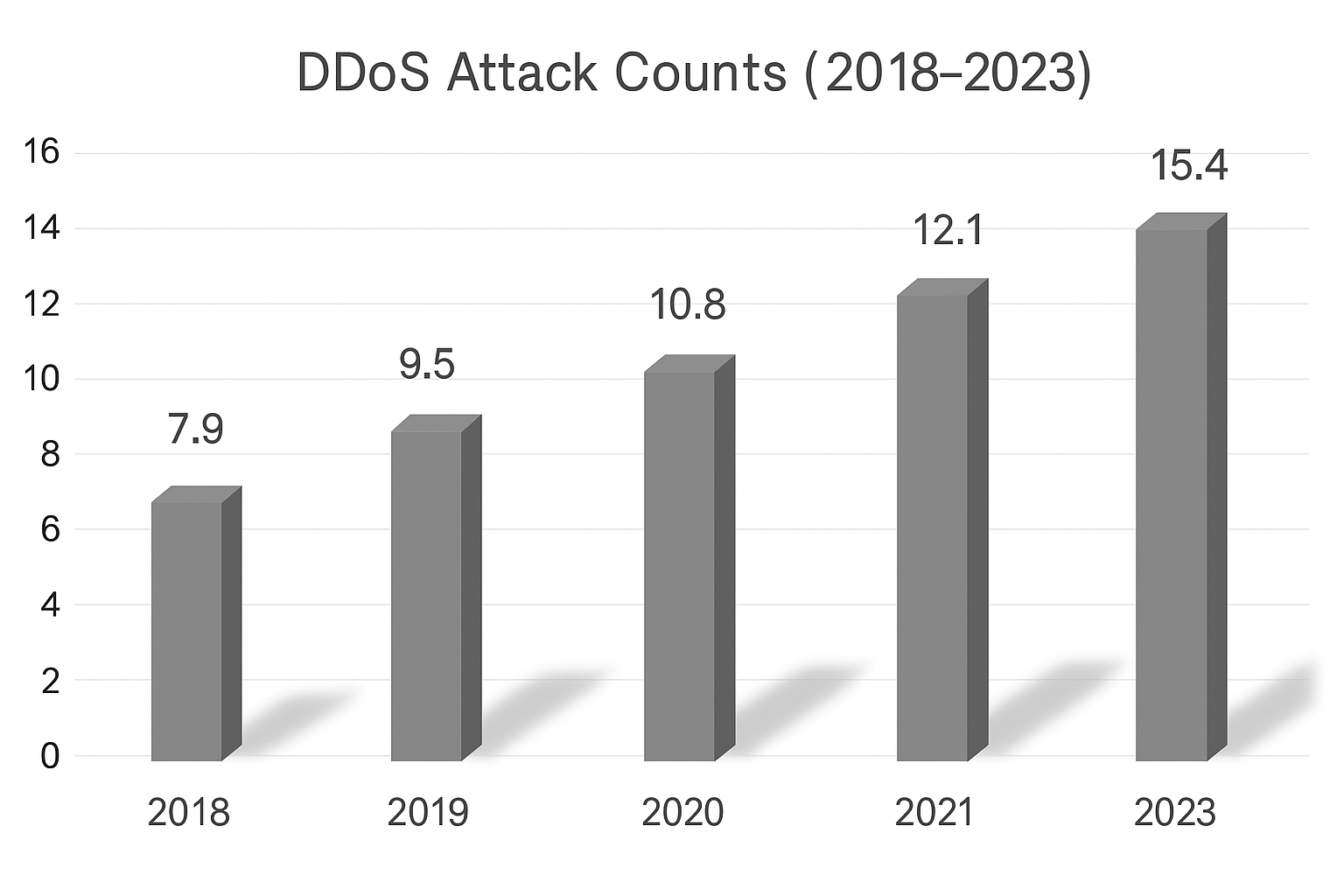}
    \caption{Number of DDoS Attack Cases 2018-2023}
    \label{fig:4-1}
\end{figure}

As shown in Fig. \ref{fig:4-1}, DDoS (Distributed Denial of Service) attacks have been continuously upgraded in recent years, presenting higher frequency, stronger attack intensity, and more complex attack methods. In response to such attacks, prior prediction of DDoS has become a key problem in the field of network security.

With its powerful feature learning and generalization capabilities, LLM provides technical support for the efficient prediction of DDoS attacks. Through in-depth modeling of massive network traffic data, large models can automatically identify potential patterns between normal and abnormal traffic, thus upgrading the security strategy from a traditional responsive mechanism to an active prediction and defense mechanism\cite{wu2025exploring}.
Feature extraction is the core aspect of the large model that plays a role in DDoS attack prediction. Using deep neural networks such as Convolutional Neural Networks (CNN), Recurrent Neural Networks (RNN), and Long Short-Term Memory Networks (LSTM), key features reflecting the state of the network, such as packet sizes, transmission intervals, communication protocols, etc., can be automatically learned from the original network stream. These models are particularly good at handling sequence data with time-dependent relationships, and thus exhibit high accuracy and robustness in anomaly detection\cite{li2024detecting}.

In addition, the model training needs to rely on large-scale datasets and combine with cross-validation and test-set evaluation to ensure that the prediction models have good generalization ability and can cope with diverse DDoS attack scenarios. Ultimately, the prediction system based on the large model can analyze network traffic in real time, and quickly issue warnings and take countermeasures when anomalous patterns are identified, so as to effectively reduce the damage caused by attacks.

In practical research, Cheon et al proposed a deep learning method based on Word2Vec for DDoS attack detection\cite{cheon2022novel}. The method significantly improves the model's ability to capture data patterns by encoding network traffic into semantic embeddings, which effectively enhances the detection performance. Meanwhile, Guastalla et al proposed an LLM detection strategy that combines Few-shot cue learning, Fine-tuning of the model, and neural network architecture. This strategy generates semantic understanding and fine-tuning optimization through LLM under the premise of using a small number of labeled samples, and ultimately achieves higher detection accuracy than traditional neural networks, which provides an innovative direction for DDoS defense in resource-constrained scenarios\cite{guastalla2023application}. The results of DDoS attack detection in practice are as follows: Listing~\ref{lst:traffic-prediction} and Tab~\ref{tab:sim_exp_results}.
\

“It's Normal Traffic” means that it is detected as normal traffic, and if it is detected as abnormal traffic, it is “Attack Traffic Detected.” The port will be blocked to prevent the attack of abnormal traffic. Port to prevent the attack of abnormal traffic.
The TPR metric represents the sensitivity, indicating that the model is more accurate; the F1-Score provides the reconciled mean value that represents the accuracy and sensitivity, and the experiments obtained high F1-Score values, so it can be considered valid.
\vspace{5pt}
\begin{lstlisting}[caption={Traffic Prediction Output}, label={lst:traffic-prediction}]
input data [165, 164, 0.012195121951219513] prediction result ['0']
It's Normal Traffic
input data [276, 276, 0.004545454545454545] prediction result ['1']
Attack Traffic detected
Mitigation Started
attack detected from port 1
Block the port 1
attack detected from port 1
Block the port 1
\end{lstlisting}

\begin{table}[h]
\centering
\caption{Simulation Experiment Results}
\label{tab:sim_exp_results}
\begin{tabular}{| c | c |}
\hline
\textbf{Metric}                & \textbf{Value}       \\ 
\hline
Precision                     & \hspace{3cm} 0.915 \hspace{3cm}  \\
\hline
Recall                        & \hspace{3cm} 0.612 \hspace{3cm}  \\
\hline
F1-Score                      & \hspace{3cm} 0.733 \hspace{3cm}  \\
\hline
Accuracy                      & \hspace{3cm} 0.747 \hspace{3cm}  \\
\hline
False Positive Rate (FPR)     & \hspace{3cm} 0.612 \hspace{3cm}  \\
\hline
True Positive Rate (TPR)      & \hspace{3cm} 0.924 \hspace{3cm}  \\
\hline
\end{tabular}
\end{table}

\subsection{Network Security Log Analysis}
\

Network security logs are an important tool for operation and maintenance, and security teams to record the operation and event information of network devices. With the expansion of network scale and the increasing complexity of the environment, the traditional manual log analysis method is difficult to meet actual needs. In recent years, the application of Large Language Model (LLM) in security log analysis has gradually become a research hotspot.
By modeling and analyzing massive logs with the help of LLM, hidden behavioral patterns and abnormal events can be effectively identified. For example, through the construction of a network log language model, abnormal access behavior can be detected, so as to warn of potential attacks timely manner. In addition, LLM can assist security teams in extracting key information from threat intelligence and automating and intellectualizing incident response.

It has been shown that deep learning and natural language processing technologies have good performance in log anomaly detection. The LogFiT model proposed by Crispin et al. is based on BERT for pre-training and fine-tuning to learn the linguistic structure of system logs\cite{almodovar2022can}. The model optimizes the normal logging pattern through two objectives: mask prediction and center-of-mass distance minimization, and combines top-k prediction accuracy and center-of-mass distance in the inference stage to determine whether the log is abnormal or not. Experimental results show that LogFiT exhibits good accuracy and robustness in the log anomaly detection task.
\subsection{Network Abnormal Traffic Detection}
\

Network anomalous traffic detection is a key link in network security analysis, aiming to discover anomalous behaviors in time by monitoring traffic changes to ensure system security. In recent years, large language models (LLMs) have been gradually introduced into this field due to their powerful feature learning and pattern modeling capabilities, and show the prospect of wide application.

LLM is able to construct language models for network packets and learn the feature distribution and behavioral patterns of normal traffic. When abnormal traffic occurs, patterns that deviate from normal behavior can be identified through model comparison, achieving efficient and accurate abnormality detection. In addition, LLM can also mine the deep dependencies in the traffic to assist analysts in more accurate threat identification.
The FlowTransformer model proposed by Manocchio et al\cite{manocchio2024flowtransformer} represents network traffic as a sequence of feature vectors (including source/destination IPs, protocol types, etc.) and uses the Transformer encoder to model its temporal dependencies to achieve classification of normal and abnormal traffic. Experiments show that this method has high accuracy and generalization ability in anomaly detection tasks.
In addition, Chai et al. proposed a detection method combining CNN and Transformer, which first extracts local features by a convolutional neural network, then models global dependencies by a Transformer encoder, and finally expresses them as abstract features by a decoder, and completes the classification with the help of SVM. The method performs well in recognizing unknown attacks and potential threats, and has good detection accuracy and practical value
\cite{chai2023ctsf}.

In this paper, an anomaly detection model that fuses a convolutional neural network (CNN) with the Transformer architecture is investigated. The model mainly includes a Transformer encoder, a convolutional feature extraction layer, and a linear support vector machine (SVM) classifier, aiming to improve the accuracy of identifying anomalous behaviors of network traffic.
The experimental data is selected from the Wednesday-working-hours section of the CIC-IDS-2017 intrusion detection dataset. In order to ensure the balance between training and testing, the dataset is divided into a training set and a testing set in the ratio of 80\% and 20\%. After the model training is completed, the first 315 samples in the test set are selected for validation, and the performance of the model is further evaluated by calculating the confusion matrix. At the same time, the model's classification effect and practical value in the anomaly detection task are comprehensively measured by combining Accuracy, Precision, Recall, and F1-score.

\begin{figure}[H]
    \centering
    \includegraphics[width=0.4\linewidth]{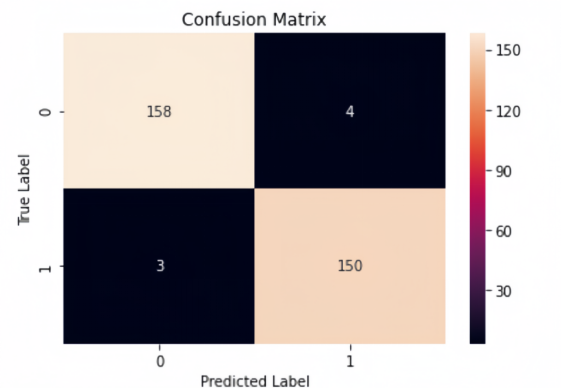}
    \caption{Confusion Matrix}
    \label{fig:4-3}
\end{figure}

After training, this paper selects the first 315 samples of the test set to compute the confusion matrix and calculates the accuracy, precision, recall, and F1 score of the model. The structure is shown in Fig. \ref{fig:4-3} and Tab~\ref{tab:simulation_results}
\begin{table}[htbp]
\centering
\caption{\large Simulated Experiment Results}
\label{tab:simulation_results}
\renewcommand{\arraystretch}{1.5}
\begin{tabular}{| @{\hspace{3cm}} c @{\hspace{3cm}} | c @{\hspace{2cm}} |} 
\hline
\textbf{\large Metric} & \textbf{\large Value} \\
\hline
\large Precision & \large 0.974 \\
\hline
\large Recall & \large 0.980 \\
\hline
\large F1-Score & \large 0.977 \\
\hline
\large Accuracy & \large 0.978 \\
\hline
\end{tabular}
\end{table}

\section{Security Vulnerability Detection}
\

Security vulnerability detection relies on an in-depth understanding of program semantics and structure, so it is crucial to select appropriate language models. Currently, commonly used models include n-gram, RNN, LSTM, and models based on the Transformer architecture, such as BERT and GPT. The Transformer model has become a mainstream solution due to its powerful semantic modeling capability.
In practice, researchers usually use large-scale pre-trained language models and fine-tune them with data from the security domain to improve the performance of the models in vulnerability detection tasks. The training process often employs a transfer learning strategy to migrate the linguistic knowledge of the general-purpose model to a specific security corpus, so as to improve the model's ability to generalize the vulnerability features\cite{bennouk2024comprehensive}.

For example, the CodeSentry framework proposed by Jones et al\cite{jones2024codesentry} is based on GPT-2, which encodes words and sub-words of C, C++, and Java code, and identifies the vulnerability types through the output vectors, which significantly reduces manual feature engineering.SmartConDetect, designed by Jeon et al., utilizes the BERT model to semantically analyze the Solidity smart contract. SmartConDetect by Jeon et al\cite{jeon2021smartcondetect}. utilizes the BERT model to semantically analyze Solidity smart contracts and realizes automated vulnerability detection through softmax classifiers, while Szabó et al. implement static analysis in combination with the GPT API, and detect sensitive snippets of front-end applications through natural language hints to effectively identify vulnerabilities such as CWE-653\cite{szabo2023new}.

The language model training method in security vulnerability detection is an important and effective technical means, making full use of deep learning and natural language processing technology, which can better discover and detect potential security vulnerabilities in the system and improve the security and stability of the system.

\section{Malicious Code Detection and Encryption Algorithm}
\subsection{Malicious Code Detection}
\

The key to malicious code detection lies in the high-quality extraction and accurate classification of features. Traditional methods mainly rely on static and dynamic analysis: static analysis extracts static features from code structure and instruction sequences, which is suitable for pre-execution detection; dynamic analysis monitors system calls, network communication, and other behaviors at runtime to reveal malicious logic. Although they complement each other in practice, there are still limitations when dealing with obfuscated code, variant samples, and complex semantic structures.

The introduction of large language models (LLMs) provides a breakthrough in malicious code analysis. With their powerful capabilities in semantic modeling, context understanding, and sequence learning, LLMs can extract deeper semantic features from complex instruction sequences and API call paths, making up for the shortcomings of traditional feature engineering\cite{li2024cobra}. At the same time, the large model can generalize the identification of variant codes, which can effectively explore the latent malicious behavior patterns and significantly improve the accuracy and adaptability of detection.
\begin{figure}[H]
    \centering
    \includegraphics[width=0.4\linewidth]{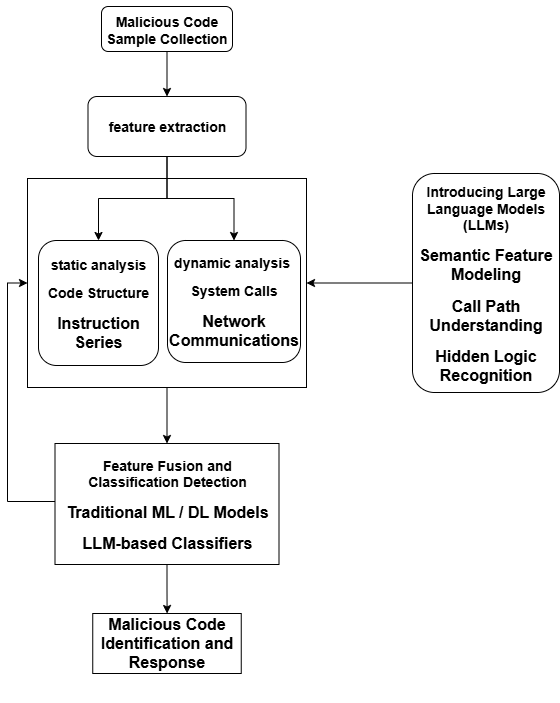}
    \caption{The Role of Large Language Modeling in the Malicious Code Detection Process}
    \label{fig:9-9}
\end{figure}
In the classification and detection stage, classification algorithms incorporating features extracted from LLMs have demonstrated superior performance. Combined with deep learning or supervised learning methods, researchers can build high-precision classifiers based on categories such as viruses, worms, Trojans, etc., to quickly recognize unknown malicious samples. In addition, LLMs can also take on the core modeling task in the end-to-end detection framework, which improves the robustness and practicality of the system in real attack and defense environments\cite{patsakis2024assessing}.
The Fig. \ref{fig:9-9} shows the core process of malicious code detection and highlights the key role of Large Language Models (LLMs) in feature extraction and semantic understanding. By fusing static and dynamic features, large models are able to model complex behavioral logic and improve detection accuracy. 
\vspace{-3ex}
\subsection{Cryptographic Algorithms}
\

With the continuous evolution of artificial intelligence technology, the application of large language models (LLMs) in symmetric encryption and asymmetric encryption algorithms has become a new hot spot in network security research. In symmetric encryption algorithms, LLMs can play a role in the following key aspects: first, the optimization of the key exchange mechanism, using its generation and prediction capabilities to assist in the establishment of a more secure key negotiation process; second, the improvement of encryption and decryption efficiency, through the identification of data patterns and contextual features, to improve the processing efficiency of the algorithm; third, the identification of vulnerabilities and risk assessment, the model can be used to analyze the history of the attack samples and the description of the protocol, to assist in discovering potential security risks. The model can analyze historical attack samples and protocol descriptions to assist in the discovery of potential security risks\cite{gupta2024enhancing}.

\begin{figure}[H]
    \centering
    \includegraphics[width=0.5\linewidth]{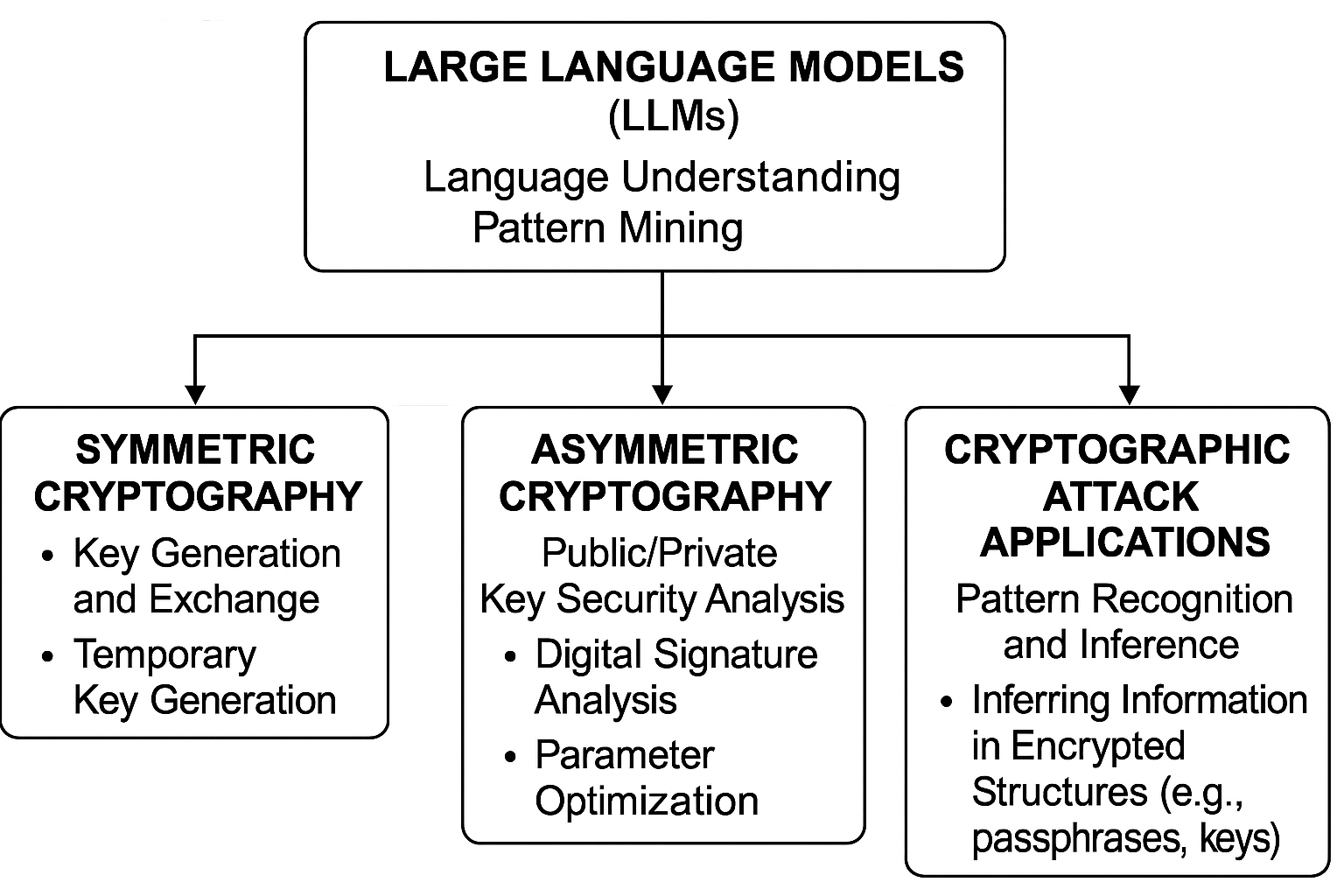}
    \caption{Large Language Modeling in Modern Cryptographic Systems}
    \label{fig:7-1}
\end{figure}

In terms of asymmetric encryption algorithms, LLMs also show application potential, mainly in key combination security assessment, digital signature logic analysis, and parameter selection optimization. With their powerful semantic understanding and reasoning capabilities, the models can be used to identify potential risks in signature structures and improve the reliability of public key infrastructure (PKI).
In addition, LLMs also show some breakthrough potential in cryptography algorithm cracking research. Compared with traditional brute-force cracking and mathematical derivation, LLMs can predict potential encryption structures by learning a large number of historical key patterns, cryptographic phrase construction laws, and user behavioral data, thus assisting in reducing the difficulty of cracking. In practical applications, this method is expected to improve efficiency in specific weak password detection and model-assisted key reconstruction scenarios. The architecture of the large Language Model for modern cryptographic systems is shown in Fig. \ref{fig:7-1}

\section{Conclusion}
\

This paper examines in detail the practical application of large language models in the field of information security technology and demonstrates the strong potential and practical value of large model technology for key issues such as malware detection, network security analysis, and security vulnerability detection. Looks forward to the future security areas where large language models may be applied, such as the field of cryptography and malicious code detection. It is shown that the use of large language models can significantly improve the detection accuracy of security threats, reduce false alarms, and speed up the threat identification and response. Experimental results validate the efficiency of these models in understanding complex security data, highlighting their superior ability to capture malicious behavior patterns as well as analyze security events.

Although the research in this paper has achieved certain results, we still need to face up to its shortcomings. In particular, the application scope and research depth of the large model are yet to be further expanded. Therefore, in our future research work, we should be committed to broadening the breadth of the study, digging deeper into the depth of the study, and continuously improving the height of the study, with a view to achieving more comprehensive and in-depth results.
Given the limitations and challenges identified in the current study, future work will focus on the following areas:
\begin{itemize}
  \item \textbf{Enhanced Model Transparency and Interpretability:} Enhance model interpretability by integrating interpretable AI mechanisms that enable security experts to better understand and trust the model's decision-making process.
  
  \item \textbf{Defending Against Adversarial Attacks:} Research and development of more robust macromodels to enhance their robustness in the face of attacks, including the design of new training methods for recognizing and confronting adversary-generated adversarial samples.
  
  \item \textbf{Enhancement of model generalization capability:} Explore ways to enhance cross-domain data applicability, e.g., using techniques such as transfer learning and multi-task learning to optimize the model's ability to generalize to different domains and types of security threats.
  
  \item \textbf{Automated Security Intelligence Collection and Analysis:} Integrate large language models to achieve fully automated threat intelligence collection, processing, and analysis, facilitating rapid sharing of security information and intelligent generation of response strategies.
  
  \item \textbf{Extended Cryptographic Applications:} Research on further applications of large language models within the field of cryptography, such as automatic generation of secure cryptographic strategies, optimization of encryption algorithms, etc.
\end{itemize}
\bibliographystyle{unsrt}
\bibliography{reference}
\end{document}